# Application of DEA in International Market Selection for the export of products from Spain

Author: Safa Elkefi 1,2,3*

[1] Industrial Engineering Department, National Engineering School of Tunis, University of Tunis El Manar, Tunisia.

[2] School of Systems and Enterprises, Stevens Institute of Technology, USA.

[3] UR-OASIS, National Engineering School of Tunis, University of Tunis El Manar, Tunisia.

**Abstract**

This article presents a Benchmarking methodology to support decision-making for international market selection (IMS). In order to do so, we will be using an output-oriented Data Envelopment Analysis (DEA) model. This methodology considers multiple variables validated with a correlation analysis. The methodology is applied to all of the products directly exported from Spain, it takes into consideration different Inputs variables and returns us the efficient and regions generating higher benefits to access international markets with the lowest costs possible.

## I- Introduction

Benchmarking methods and performance analysis are essential for decision-making according to Armado, 2005[1], especially when it comes to a multi-target decision with several key variables, constraints with several input (Inputs) and output (Outputs) variables.

The idea is to guarantee both economy and efficiency without affecting the results.

We will explain later how to make a better choice by keeping the same quality of productivity especially when it comes to quantifiable variables and not only to approaches and methods to follow which are interpretable by the human being.

Productivity is the ratio between the Output (Y) and the Input (X) such as:

$$\text{Productivity} = \frac{Y}{X} \qquad (1)$$

Solving a productivity optimizing problem is finding possible areas where we can act on the input and output variables to achieve better efficiency or better use of resources.

This may be done by two approaches: Input and Output oriented methods.

The Output orientation can be defined as the amount with which we can increase proportionally the production without changing input quantities used. The Input orientation however is to decide how much we can proportionally reduce the input quantities without changing the quantities produced [1].

It is easy to decide when it is a single Input and Output. However, when they are more than one, a single ratio is not enough. This makes us in need for a more complex Benchmarking model such as DEA or the Data Envelopment Analysis method.

Under the digitalization of its processes, a company specialized in international sales has planned a full digitalization of its activity in order to optimize its resources. Among the main processes optimized, we can find market selection automatization, automatically selecting the countries that a client can approach using the help of a well-established strategy. For this, we used DEA method (Data Envelopment Analysis).

## II- DEA method
### 1) Applications of DEA

Data Envelopment Analysis (DEA) is a non-parametric linear programming technique that measures relative performance of decision-making units (DMUs) as a function of the values of the selected input and output variables by avoiding weight assignment and normalization of variable values on a homogeneous scale [2]. Therefore, these issues include model orientation and selection or definition of inputs, outputs, and Decision-Making Units (DMUs).

This technique makes it appliable in many domains such as banks, mutual funds, police stations, hospitals, tax authorities, military defense bases, insurance companies, schools, libraries, and university departments. It can be considered as a tool to solve multi-criteria evaluation problem in which DMUs are alternatives and each is represented by its performance according to several criteria classified in the DEA input and output model [3]. Currently, some variables may be strategic in an integrated management system such as institutional support from governments to protect local businesses, geographic proximity, cultural similarity, and other key variables [4].

In addition, interesting contributions show the relevance of the DEA in International Market Selection approaches (IMS) [16]. Various studies have applied the approach by using measures such as customer image, average tax rates in target markets, number of foreign competitors, number of local manufacturers, marketing costs, sales prices, and estimated sales volume [6]. Governments and companies are also making great efforts to develop customized softwares based on DEA techniques to recognize opportunities in international markets in order to develop alternatives for economic growth [7], [16].

For each company seeking to increase its involvement in international markets through exported products, we must identify the most optimal solution given obstacles, criteria and important institutional mechanisms that can affect an export process. DEA models allow companies to take into consideration many factors to identify the most efficient target countries, such as measuring and evaluating the potential of international shipping line flows and their effectiveness in containerized freight. On this basis, in a real application of DEA for IMS, decision takers must identify the appropriate Inputs and Outputs according to their needs and preferences [8] which was proposed by Charnes, Cooper and Rhodes in 1978 [9].

The DEA is also called border analysis because the performance of a unit is assessed by comparing it with the best performing units in a sample representing an efficiency border. In addition, if the unit is not at the limit of its effectiveness, it is considered ineffective. Therefore, this multi-criteria method can be applied in our case because it can manage several inputs and outputs as opposed to other quantitative techniques such as ratio analysis or regression methods.

### 2) Presentation of the model and the data

In this section, we focus on the application of DEA in the Selection of International Markets.

We start by defining the variables of the model. The first variable to be defined is the list of products to study their markets. We consider the distribution of the European Product Commission according to the HS (Harmonized Systems) code with a depth coding of only 6 digits for a compromise between ensuring data availability and entering more accurate identification. This way, we get 21 sections that contain 97 chapters with a total of 617 product families. For each of the products, we want to identify the target market.

After identifying the list of products, we collect all the variables that can be used during decision-making to enrich the model. During learning as much possible of the export operations process, the idea of identifying a domain of variables is essential. We have therefore considered moving beyond the ordinary framework of selecting target countries on the basis of their purchasing power and import history and involving other variables that may have an influence on these operations.

Given the importance of the reliability of data sources, we thought about narrowing our search scope to a list of recognized databases such as the World Bank, the International Monetary Fund (IMF), the Hofstede Center, ITC and TradeMap. The set of variables retained from these databases includes 35 variables for the set of 232 countries according to ITC's world distribution. The ITC (International Trade Center) country list is the Benchmark for data collection. The criteria defined by [10] and [11] were considered to determine Inputs and Outputs, which include variables at the country and consumer level related to economic and market development, product acceptance, cultural and geographical distances, logistics, policies facilitating the creation of activities and the movement of goods among others.

For the variables in the database, a correlation analysis between the input and output variables was also performed to ensure that each variable provided information different from the other variables in the model, thus increasing the contrast between the countries evaluated. This analysis facilitates the identification of variables that are convenient to store or not and has been performed for all chapter datasets at 4-digit depth.

The inputs count M well correlated variables such as M is equal to 10. They define the cost of an export operation from a country to another such as the cultural gap's indexes and the logistical issues, procedures, and documents. Tables 1 below contains the variables of Inputs and Outputs respectively.

| Variable | Input/Output |
|---|---|
| Average distance of importing countries in km | Input |
| Logistics performance index | |
| Index of ease of doing business | Input |
| GDP growth 2014-2018GDP per capta in 2018 | Input |
| Customs procedures load | Input |
| Export lead time | Input |
| Cultural Gap compared to Spain : long-term orientation | Input |
| Cultural Gap compared to Spain : Avoid uncertainty | Input |
| Cultural Gap compared to Spain : Power Distance | Input |
| Imported value in 2018 | Output |
| Trade balance | Output |
| Quantity imported in 2018 | Output |
| Annual growth in value 2014-2018 | Output |
| Annual quantity growth 2014-2018 | Output |
| Annual growth in value 2017-2018 | Output |
| Annual quantity growth 2017-2018 | Output |
| Expected Commodity Imports between 2018-2020 | Output |
| Share in world exports | Output |
| Concentration of importing countries | Output |

**Table 1: DEA Model Decision Variables**

Once the inputs and outputs (S variables = 10) have been set and according to Friedman and Sinuany-Stern, 1998 [12], the number of DMUs must be at least equal to (M + S) × 3. In this case, it is (10 + 10) × 3 = 60 countries. And this is consistent with the 233 countries in the dataset. In this context, the condition established by Allen et al, 2001 [13] would also be met since the number of DMUs in DEA models should be at least equal to 2M × S, which in this case is 2 (10) × 10 = 200 countries. The DMUs number used for international market selection lets us aim for a higher level of segregation and gives the model more degrees of freedom.

3) **DEA Mathematical Modeling Methodology**

The BCC models of Banker, Charnes and Cooper [14] are used to select international markets for the product we wish to export from Spain based on the nature of the inputs and outputs described in Tables 1 and 2.

The DEA thus makes it possible to identify when a company is in a position to compete in international markets and to recognize market export opportunities [15].

The Inputs considered can be interpreted as the costs and efforts required to access an international market. Inputs with smaller values will be useful for business performance. Therefore, the Outputs considered can be interpreted as the benefits offered by a target country, and Outputs with higher values will be more practical for business performance.

In this case, an Output-oriented model is used because the decision-maker is interested in finding a country where the benefits of exporting his product are maximized given certain costs and efforts. To solve this problem, we use the DEA method with output orientation, which consists of determining efficiency Benchmarks (reference DMUs) and locating all other decision units in relation to these Benchmarks. It uses data wrapping: the units on the envelope (or empirical productivity frontier) are therefore the reference points. A distance from other units at this production boundary is a measure of their inefficiency. Maximizing the efficiency of each $DMU_k$, of the BCC Output-oriented model, is equivalent to maximizing its DEA ratio of equation (2) by remaining within the achievable range of constraint modelled solutions (3) with a positive weighting of Inputs (4) and Outputs (5):

$$Max: \theta(k) = \frac{\sum_{t=1}^{S} U(r)Y(r,k)}{\sum_{i=1}^{M} V(i)X(i,k)} \quad (2)$$

$$\text{S.To:} \quad \theta(j) \leq 1, \forall\, j=1,..,N \quad (3)$$

$$U(r) \geq 0, \forall\, r=1,..,S \quad (4)$$

$$V(i) \geq 0, \forall\, i=1,..,M \quad (5)$$

With:

K: The Benchmark

$\theta(k)$: Ratio of DEA "for the DMU(k)

Y(r,j): Quantity of Output r for the DMU(j)

X(i,j): Quantity of Input i for the DMU(j)

V(i): Weight of the Input i

U(r): Weight of the Output r

S: Number of Outputs

M: Number of Inputs

N: Number of Decision-Making Units (DMU)

An assessment of the productivity of a Benchmark k country or a decision variable of index k (DMU(k), k=1, .. ,233), will therefore be none other than the assessment of the efficiency score $\theta(k)$ of the equation (2) that we need to maximize while remaining in the realizable domain. For an output-oriented BCC model, $\theta$ also represents the radial expansion of the Outputs for a maximum level of inputs. Therefore, if $\theta > 1$, which is equivalent to saying that there is a combination of outputs from the other DMUs that is greater than that of the assessed DMUP, then it is classified as an ineffective DMU. In this case, the Inputs considered represent costs and efforts necessary to access an international market. Therefore, Inputs with smaller values will be more useful for business performance and Outputs can be interpreted as the benefits offered by a destination country so Outputs with higher values will be more practical for performance. It is in our interest to minimize costs and maximize profits while remaining in the achievable domain of solutions.

This model is, algebraically, a solution to the following problem [6]:

$$Max: \theta(k) = \frac{\sum_{r=1}^{10} U(r)Y(r,k)}{\sum_{i=1}^{10} V(i)X(i,k)} \quad (6)$$

$$\text{S.to: } \sum_{j=1}^{233} \lambda(j)Y(r,j) \geq \theta(k)Y(r,k), \forall\ r=1,...,10 \quad (7)$$

$$\sum_{j=1}^{233} \lambda(j)X(i,j) \leq X(i,k), \forall\ i=1,...,10 \quad (8)$$

$$\sum_{j=1}^{233} \lambda(j) = 1 \quad (9)$$

$$\lambda(j) \geq 0, \forall\ j=1,...,233 \quad (10)$$

### 4) Solving the problem and results

To load the necessary data and solve this problem, we have developed a DEA algorithm with R language under the packages *lpsolveAPI, Benchmarking* and *ucminf*. With these packages, we can apply functions that make Benchmarking for each product easy. Once the data is selected, we apply through the algorithm the functions that allow us to solve the problem by analyzing data development and calculating the DEA ratios then select effective Benchmarks, whose efficiency is equal to 1.

Then we make another filter to order the countries and take the first 5. This filter consists of considering only the variable *Quantity imported in 2018*, weighting it by the sum of the quantities imported in 2018 for the product by all the countries in the world and then ordering the countries already selected according to the increasing percentage of their imports. This approach was used to select the target countries for the 617 products.

### III- Conclusion

This paper investigated the Data Envelopment Analysis Problem which is known as DEA Benchmarking method. This method has many applications and offers a good explanation of Decision-Making Unit and its efficiency compared to others. The BBC output-oriented model in this case was applied to evaluate the efficiency of international markets to export all products to all the world from Spain and to identify the best countries having the highest benefits from the lowest costs, taking into consideration multiple criteria, measures, and indexes. According to the preferences of the decision-maker and for other applications, DMU selection criteria may change, and some inputs and outputs can be customized.